\documentclass[12pt]{article}
\usepackage{amsmath}
\usepackage{amssymb}
\usepackage{bm}
\newcommand{\be}{\begin{equation}}
\newcommand{\ee}{\end{equation}}
\newcommand{\bea}{\begin{eqnarray}}
\newcommand{\eea}{\end{eqnarray}}

\newcommand{\mbss}[1]{_{\mbox{\scriptsize #1}}}
\newcommand{\mbts}[1]{_{\mbox{\tiny #1}}}
\newcommand{\mbsu}[1]{\mbox{\scriptsize #1}}

\newcommand{\ds}{\displaystyle}

\newcommand{\vk}{\varkappa}
\newcommand{\mbold}[1]{\mbox{\boldmath $#1$}}
\newcommand{\bnabla}{\mbold{\nabla}}
\makeatletter
\renewcommand{\section}{\@startsection{section}{1}{0pt}%
{-3.5ex plus -1ex minus -.2ex}{2.3ex plus.2ex}%
{\normalsize\bf}}
\makeatother
\begin{document}
\title{%
\large\bf
 DENSITY MATRIX FUNCTIONAL THEORY\\
 WITH ACCOUNT OF PAIRING CORRELATIONS}
\author{%
S. Krewald,$^1\;$
V. B. Soubbotin,$^2\;$
V. I. Tselyaev,$^2\,$ and
X. Vi\~nas$\hphantom{,}$$^3$\\$\vphantom{,}$\\
{\it \normalsize
 $^1$
 Institut f\"ur Kernphysik,
 Forschungszentrum J\"ulich,
 D-52425 J\"ulich, Germany}$\vphantom{,}$\\
{\it \normalsize
 $^2$
 Nuclear Physics Department,
 V. A. Fock Institute of Physics,}\\
{\it \normalsize
 St. Petersburg State University, 198504,
 St. Petersburg, Russia}$\vphantom{,}$\\
{\it \normalsize
 $^3$
 Departament d'Estructura i Constituents de la Mat\`eria,}\\
{\it \normalsize
 Facultat de F\'{\i}sica, Universitat de Barcelona,}\\
{\it \normalsize
 Diagonal 645 E-08028 Barcelona, Spain}}
\date{December 6, 2004}
\maketitle
\begin{abstract}
The extension of the density functional theory (DFT) to include
pairing correlations without formal violation of the
particle-number conservation condition is described.
This version of the theory can be considered as a foundation
of the application of existing DFT plus pairing approaches to atoms,
molecules, ultracooled and magnetically trapped atomic Fermi gases,
and atomic nuclei where the number of particles is
exactly conserved. The connection with the Hartree-Fock-Bogoliubov
theory is discussed. The method of the quasilocal reduction of
the nonlocal theory is described. This quasilocal reduction allows
to obtain equations of motion which are much more simple for the numerical
solution than the equations corresponding to the nonlocal case.
\vspace{2em}
\begin{flushleft}
\end{flushleft}
\end{abstract}
\newpage
\section{INTRODUCTION}

Although both the theory of superconductivity and
the density functional theory (DFT) have a long history,
models which take into account pairing correlations within
the framework of the DFT have appeared not so long ago.
The initial version of the DFT developed in the early papers of
Hohenberg, Kohn, and Sham \cite{HK,KS} did not include the
pairing correlations explicitly. The first generalization
of the DFT in this direction was developed in Ref.~\cite{OGK}
for superconductors. However, the essential feature of this theory
is the nonconservation of the number of particles in the
superconducting Fermi system. Consequently, the application
of the DFT for superconductors to atoms, molecules,
ultracooled and magnetically trapped atomic Fermi gases, and
atomic nuclei, where the pairing correlations may be important
but the number of particles is exactly conserved, requires
an additional foundation. In fact, the same question emerges
in connection with the DFT plus pairing approaches based, for example,
on the theory of finite Fermi systems of Migdal
(see \cite{FTTZ} and references therein) or on the local density
approximation (Ref.~\cite{B}). So the first goal of the present
paper is a rigorous formulation of the extended version
of the DFT taking into account the pairing correlations
under the condition of particle-number conservation.

The second goal is to extend the DFT to the case of
a functional dependence on the total nonlocal single-particle
density matrix (DM). Such an extension is especially important
in applications to nuclei because it allows to introduce
in a natural way the kinetic-energy and the spin densities
dependence of the energy functional. That kind of dependence
leads to the appearance of a radial-dependent effective mass
and a spin-orbit potential which are essential components of the
nuclear structure models. Let us note that the original version
of the DFT can be classed as a local theory because in the
papers \cite{HK,KS} the energy functional only depends on the
local particle density.
One of the possible ways of a nonlocal extension of the DFT was
considered in Ref.~\cite{Gilb}. However, this method,
which can be referred to as a straightforward extension, faces
with serious difficulties related with the equations of motion
and their physical interpretation.
Another method has been recently developed in Ref.~\cite{STV} to
extend the DFT by considering an
energy functional which depends on a Slater-determinant DM.
This approach leads to the quasilocal density functional theory and
allows to avoid the difficulties
arising from the method described in \cite{Gilb} but, in general, the
resulting DM is not the exact DM because only its diagonal part is the
quantity which coincides  with the exact local particle density
of the interacting fermion system.

In the present paper we will show that the inclusion of the pairing
within the framework of the extended DFT
is enough to obtain the exact
total nonlocal DM on the one hand, and to avoid the difficulties
encountered in the equations of motion of Ref.~\cite{Gilb}
on the other hand.
Although we develop our formalism
in the particular case of atomic nucleus, the main results
and conclusions can also be applied to any Fermi system with a fixed
number of particles.
The paper is
organized as follows. In Sec.~\ref{sec2} the extended density matrix (EDM)
formalism is revised.
In Sec.~\ref{sec3} the DFT is extended in order to include
the pairing correlations.
In Sec.~\ref{sec4} the reduction to
an extended quasilocal density functional theory is performed.
The conclusions are given in the last section.
In the Appendices some mathematical details and comments are presented.

\section{EXTENDED DENSITY MATRIX FORMALISM \label{sec2}}

Despite the formalism of the density matrix extended to include
pairing correlations in the ground state is well known (see,
e.~g., Refs.~\cite{RS,edm}), we shall draw some basic definitions
which are necessary for the further analysis.
Let $\Psi$ be some arbitrary antisymmetrized many-fermion
wave function. In the general case $\Psi$ is assumed to be normalized,
but it is not supposed to be an eigenfunction of the particle-number
operator. Thus the normal ($\rho$) and the anomalous ($\kappa$)
density matrices in the state $\Psi$ are defined through the
following expectation values:
\be
\rho (x,x') = \langle \Psi |\, a^{\dag} (x')\, a (x)\,
| \Psi \rangle \,,
\label{rho}
\ee
\be
\kappa (x,x') = \langle \Psi |\, a (x')\, a (x)\,
| \Psi \rangle\,,\quad
\kappa^* (x,x') = \langle \Psi |\,
a^{\dag} (x)\, a^{\dag} (x')\, | \Psi \rangle\,,
\label{kappa}
\ee
where $a^{\dag} (x)$ and $a (x)$
are creation and annihilation operators of particles in the
coordinate representation of the usual single-particle space.
In the case of atomic nuclei,
symbol $x = \{{\mbold r}, \sigma, q \}$
includes the spatial coordinate $\mbold r$ and the
spin projection $\sigma$ variables as well as the index $q=n,p$
indicating the nucleon type (neutrons and protons).

It is important to note that for Fermi systems,
which are considered here,
the ground state (GS) is described by a wave function
$\Psi = \Psi_{\mbss{GS}}$ with a fixed number of particles.
Therefore, in the GS the anomalous DM vanishes:
\be
\kappa_{\mbss{GS}} (x,x') = 0\,.
\label{kGS}
\ee
However, in many physical problems one can construct auxiliary
quantities, which have the sense of an anomalous DM but nevertheless
take nonzero values in the GS even if the Eq.~(\ref{kGS}) is
fulfilled.

Let us suppose that the DM $\rho$ defined by Eq.~(\ref{rho})
is given for some wave function $\Psi$ (in particular, it
may be the exact GS wave function $\Psi_{\mbss{GS}}$). If $\Psi$
is time-reversal invariant, we can introduce the canonical basis (CB)
$\{ \phi_{\lambda}(x) \}$,
where the single-particle multiindex $\lambda$ contains the
sign of the spin projection $s$ and the set of the remaining
quantum numbers $\{c\}$ ($\lambda = \{ c,s \}$),
so that the following expansion is fulfilled
\be
\rho (x,x') = \sum_{c,s} v^2_c\,
\phi^{\vphantom{*}}_{c,s}(x)\, \phi^*_{c,s}(x')\,,
\quad 0 \leqslant v_c \leqslant 1\,.
\label{rhoe}
\ee
In the case of real Fermi systems, this DM is not idempotent
(i.~e. $\rho^2 \ne \rho$). This is due to the fact that all
or almost all the eigenvalues of $\rho$,
defined by the equation $\int dx'\,\rho(x,x')\,
\phi^{\vphantom{*}}_{c,s}(x') =
v^2_c\,\phi^{\vphantom{*}}_{c,s}(x)$
(hereinafter $\int dx$ means the space integral over ${\mbold r}$
and the sum over $\sigma$ and $q$ indices),
lie in the interval $0 < v^2_c <1$.
If one uses, as it was done in the theory developed
by Gilbert in Ref.~\cite{Gilb}, an energy functional
$E_{\mbss{G}} [\rho]$ depending only on such
{\sl nonlocal} and non-idempotent DM,
in the resulting equations of motion all
partially occupied natural spin orbitals
(i.~e. functions $\phi^{\vphantom{*}}_{c,s}(x)$
for which $0 < v^2_c <1$)
are eigenfunctions of the single-particle pseudo-Hamiltonian
$h_{\mbss{G}} (x,x') = \delta E_{\mbss{G}}[\rho]/ \delta \rho (x',x)$
with the same eigenvalue (see Ref.~\cite{Gilb} for details).
This fact leads to difficulties in the physical interpretation
and in the mathematical foundation of the theory developed by Gilbert.
To avoid this problem,
first of all we shall define an extended density matrix (EDM)
$\cal{R}$ which has to be idempotent (${\cal R}^2 = {\cal R}$)
on the one hand and has to contain a given DM $\rho$
as a block on the other hand. An EDM means a
DM which is defined in a space being the doubled
usual single-particle space.

Let $\{ \psi_{\lambda ; \eta}(x;\chi) \}$
be an arbitrary set of basis functions in this extended space
spanned by the coordinates $\{ x, \chi \}$,
where $\chi = \pm 1$ and $\eta = \pm 1$ are additional indices
introduced for denoting the different components of the single-particle
functions. The meaning of these indices will be specified in the following
(in particular, the index $\chi$ indicates the upper and lower
components of the functions $\psi_{\lambda ; \eta}(x;\chi)$
according to notations of Ref.~\cite{edm}).
The usual conditions of
orthonormality and completeness are supposed to be satisfied:
\bea
\sum_{\chi} \int dx\,
\psi^*_{\vphantom{\lambda' ; \eta'}\lambda ; \eta}(x;\chi)\,
\psi^{\vphantom{*}}_{\lambda' ; \eta'}(x;\chi) &=&
\delta_{\eta,\eta'}\, \delta_{\lambda,\lambda'}\,,
\label{orth}\\
\sum_{\lambda,\eta}
\psi^*_{\vphantom{\lambda' ; \eta'}\lambda ; \eta}(x;\chi)\,
\psi^{\vphantom{*}}_{\vphantom{\lambda' ; \eta'}\lambda ; \eta}
(x';\chi') &=&
\delta_{\chi,\chi'}\, \delta (x,x')\,,
\label{comp}
\eea
where
$\delta (x,x') = \delta({\mbold r}-{\mbold r}') \,
\delta_{\sigma,\sigma'}\, \delta_{q,q'}$.
In addition we also assume that the functions
$\psi_{\lambda ; \eta}(x;\chi)$ satisfy the condition:
\be
\psi^{\vphantom{*}}_{\lambda ; \eta}(x;\chi) =
\psi^*_{\lambda ; -\eta}(x;-\chi)\,.
\label{conj}
\ee

From the conditions (\ref{orth}) and (\ref{comp}) it follows
that the functions $\psi_{\lambda ; \eta}(x;\chi)$ form
a unitary matrix in the extended space defined previously.
(Strictly speaking, they form a unitary operator.
The use of the term {\sl matrix} implies that the configuration
space is discretized and restricted by a finite number of points.)
If the condition (\ref{conj}) is also fulfilled,
the Bloch-Messiah theorem (see Refs.~\cite{RS,BM})
can be applied to this matrix.
In order to reformulate this theorem in coordinate representation let us
first introduce a complete set of orthonormal functions
$\{ \tilde{\phi}_{\lambda}(x) \}$. Notice that these functions form a
unitary matrix $D$ in the notation of Ref.~\cite{RS} according to the
rule: $D_{\,i\,k} = \tilde{\phi}_{\,\lambda_{\scriptstyle k}}
(x_{\vphantom{\lambda}_{\scriptstyle i}})$. Second, let us divide the set
of the single-particle indices $\lambda$ into three subsets: two sets of
conjugate indices $p$ and $\bar{p}$ which represent``paired''
states and the set of the indices $b$ corresponding to ``blocked'' states,
i.~e.: $\{ \lambda \} = \{ p \} \cup \{ \bar{p} \} \cup \{ b \}$.
Let, further, $\tilde{v}_{\lambda}$ and $\tilde{u}_{\lambda}$
be real non-negative numbers which satisfy the following conditions:
$\,\tilde{u}^{\vphantom{2}}_{\lambda} =
\sqrt{1 - \tilde{v}^2_{\lambda} \vphantom{V^A}}\,$,
$\,\tilde{v}_{p} = \tilde{v}_{\bar{p}}\,$,
$\,0 < \tilde{v}_{p} < 1\,$,
$\,\tilde{v}^2_{b} = \tilde{v}^{\vphantom{2}}_{b}\,$.
According to the Bloch-Messiah theorem
the functions $\psi_{\lambda ; \eta}(x;\chi)$ can be
represented in the following form
\be
\psi_{\lambda ; +}(x;\chi) = \sum_{\lambda'}
C^{\vphantom{*}}_{\lambda'\lambda}\,
\check{\psi}_{\lambda' ; +}(x;\chi)\,, \qquad
\psi_{\lambda ; -}(x;\chi) = \sum_{\lambda'}
C^*_{\lambda'\lambda}\,\check{\psi}_{\lambda' ; -}(x;\chi)\,,
\label{unitr}
\ee
where $C^{\vphantom{*}}_{\lambda'\lambda}$ is a unitary matrix, and
the functions $\check{\psi}_{\lambda ; \eta}(x;\chi)$ have the form:
\be
\left.
\begin{array}{rclrclrcl}
\check{\psi}_{p\,;+}(x;+)\!\!\! &=& \!\!\!\hphantom{-}
\tilde{u}^{\vphantom{*}}_{p}\,
\tilde{\phi}^{\vphantom{*}}_{p}(x)\,,\quad&
\check{\psi}_{\bar{p}\,;+}(x;+)\!\!\! &=& \!\!
\tilde{u}^{\vphantom{*}}_{p}\,
\tilde{\phi}^{\vphantom{*}}_{\bar{p}}(x)\,,\quad&
\check{\psi}_{b\,;+}(x;+)\!\!\! &=& \!\!
\tilde{u}^{\vphantom{*}}_{b}\,
\tilde{\phi}^{\vphantom{*}}_{b}(x)\,,\\
\check{\psi}_{p\,;+}(x;-)\!\!\! &=& \!\!\!
-\tilde{v}^{\vphantom{*}}_{p}\,
\tilde{\phi}^*_{\bar{p}}(x)\,,\quad&
\check{\psi}_{\bar{p}\,;+}(x;-)\!\!\! &=& \!\!
\tilde{v}^{\vphantom{*}}_{p}\,
\tilde{\phi}^*_{p}(x)\,,\quad&
\check{\psi}_{b\,;+}(x;-)\!\!\! &=& \!\!
\tilde{v}^{\vphantom{*}}_{b}\,
\tilde{\phi}^*_{b}(x)\,,\\
\check{\psi}_{p\,;-}(x;+)\!\!\! &=& \!\!\!
-\tilde{v}^{\vphantom{*}}_{p}\,
\tilde{\phi}^{\vphantom{*}}_{\bar{p}}(x)\,,\quad&
\check{\psi}_{\bar{p}\,;-}(x;+)\!\!\! &=& \!\!
\tilde{v}^{\vphantom{*}}_{p}\,
\tilde{\phi}^{\vphantom{*}}_{p}(x)\,,\quad&
\check{\psi}_{b\,;-}(x;+)\!\!\! &=& \!\!
\tilde{v}^{\vphantom{*}}_{b}\,
\tilde{\phi}^{\vphantom{*}}_{b}(x)\,,\\
\check{\psi}_{p\,;-}(x;-)\!\!\! &=& \!\!\!\hphantom{-}
\tilde{u}^{\vphantom{*}}_{p}\,
\tilde{\phi}^*_{p}(x)\,,\quad&
\check{\psi}_{\bar{p}\,;-}(x;-)\!\!\! &=& \!\!
\tilde{u}^{\vphantom{*}}_{p}\,
\tilde{\phi}^*_{\bar{p}}(x)\,,\quad&
\check{\psi}_{b\,;-}(x;-)\!\!\! &=& \!\!
\tilde{u}^{\vphantom{*}}_{b}\,
\tilde{\phi}^*_{b}(x)\,.\\
\end{array}
\right\}
\label{cbc}
\ee

Let us now define the EDM $\cal R$ in terms
of the arbitrary set of functions
$\{ \psi_{\lambda ; \eta}(x;\chi) \}$ introduced above by the formula
\be
{\cal R}(x,\chi\, ;\, x',\chi') =
\sum_{\lambda}
\psi^{\vphantom{*}}_{\lambda;-}(x;\chi)\,
\psi^*_{\lambda;-}(x';\chi')\,.
\label{defedm}
\ee
Using Eqs. (\ref{orth}) -- (\ref{conj}) it can be easily shown that
the following equalities are fulfilled:
\be
{\cal R}^2 = {\cal R}\,, \qquad {\cal R}^{\dag} = {\cal R}\,,
\label{rprop1}
\ee
\be
{\cal R}(x,\chi\, ;\, x',\chi') =
\delta_{\chi,\chi'}\, \delta (x,\,x') -
{\cal R}(x',-\chi'\, ;\, x,-\chi)\,.
\label{rprop2}
\ee

Let us introduce notations for the blocks of the EDM taking
into account the properties (\ref{rprop1}) -- (\ref{rprop2}):
\be
\left.
\begin{array}{ll}
{\cal R}(x,+\, ;\, x',+) = \tilde{\rho} (x,x')\,, &
{\cal R}(x,+\, ;\, x',-) = \tilde{\kappa} (x,x')\,, \\
{\cal R}(x,-\, ;\, x',+) = - \tilde{\kappa}^* (x,x')\,,\quad &
{\cal R}(x,-\, ;\, x',-) = \delta(x,x') - \tilde{\rho}^* (x,x')\,, \\
\end{array}
\right\}
\label{rnot}
\ee
or in the matrix representation:
\be
{\cal R} = \left(
\begin{array}{cc}
\tilde{\rho} & \tilde{\kappa} \\
- \tilde{\kappa}^* &\quad 1 - \tilde{\rho}^* \\
\end{array}
\right)\,.
\label{rmatr}
\ee
As it will be clear in the following
(see Eqs.~(\ref{trho}) -- (\ref{tkappa}) below),
the blocks $\tilde{\rho}$ and $\tilde{\kappa}$ play the role
of the normal and anomalous density matrices in some
(quasi-particle vacuum) state.

Substituting Eqs. (\ref{unitr}) and (\ref{cbc}) into
Eq.~(\ref{defedm}), we obtain:
\be
\tilde{\rho} (x,x') = \sum_{\lambda} \tilde{v}^2_{\lambda}\,
\tilde{\phi}^{\vphantom{*}}_{\lambda}(x)\,
\tilde{\phi}^*_{\lambda}(x')\,,
\label{trhoe}
\ee
where the sum over $\lambda$ implies the sum over the three sets
of indices $p$, $\bar{p}$, and $b$. Thus, the functions
$\tilde{\phi}^{\vphantom{*}}_{\lambda}(x)$ form the CB in which
$\tilde{\rho} (x,x')$ is diagonal.
So in the following we shall refer to Eqs.~(\ref{cbc}) as the CB
representation of the functions $\psi_{\lambda ; \eta}(x;\chi)$.

Comparing Eqs. (\ref{rhoe}) and (\ref{trhoe}) we see that the
expansion (\ref{rhoe}) is a particular case of (\ref{trhoe}).
Indeed, setting the number of blocked occupied states in
(\ref{trhoe}) to be even or to be equal to zero,
one can choose $\tilde{v}_{\lambda} = v_{\lambda}$,
$\tilde{\phi}^{\vphantom{*}}_{\lambda}(x) =
\phi^{\vphantom{*}}_{\lambda}(x)$, which leads to the
coincidence of the right-hand sides of Eqs.~(\ref{rhoe})
and (\ref{trhoe}). So, there exists a variety of
sets of the functions $\psi_{\lambda ; \eta}(x;\chi)$,
which enter into the definition (\ref{defedm}) and
differ by an arbitrary unitary transformation
of the type Eq.~(\ref{unitr}), such
that the following equalities are fulfilled:
\be
{\cal R}(x,+\, ;\, x',+) = \tilde{\rho} (x,x') = \rho (x,x')\,.
\label{rprop3}
\ee
Due to the fact that Eq.~(\ref{rhoe})
is fulfilled for any physically meaningful DM $\rho$,
actually we have proved that for an arbitrary {\sl nonlocal}
DM $\rho (x,x')$, corresponding to some interacting
time-reversal invariant fermion system,
we can construct an EDM $\cal R$ which satisfies the conditions
(\ref{rprop1}) -- (\ref{rprop2}) and which is related with
the DM $\rho$ by the formula (\ref{rprop3}).

It is important to note that, if the normal DM $\rho$ is produced
by some wave function $\Psi$ according to Eq.~(\ref{rho}),
the anomalous DM $\kappa$ produced by the same
wave function $\Psi$ according to Eqs.~(\ref{kappa})
does not coincide, in general, with
the quantity $\tilde{\kappa}$ defined as a block of the EDM
by Eqs.~(\ref{rnot}), even if
Eqs.~(\ref{rprop3}) are fulfilled.
In particular, if $\Psi = \Psi_{\mbss{GS}}$ then, as it was
mentioned above, $\kappa = \kappa_{\mbss{GS}} = 0$,
but for the interacting system $\tilde{\kappa} \ne 0$.
On the other hand, if $\Psi$ is a quasiparticle-vacuum
wave function (see below), then the equality
$\kappa = \tilde{\kappa}$ is fulfilled.

In order to reproduce an arbitrarily given DM $\rho (x,x')$
as a block of the EDM $\cal R$
we have started from an arbitrary complete set of basis functions
$\{ \psi_{\lambda ; \eta}(x;\chi) \}$ which satisfies the conditions
(\ref{orth}) -- (\ref{conj}).
It is useful to do it in a different way,
looking at this problem from another
more traditional point of view.
To this end, let us introduce the
creation and annihilation operators of the quasiparticles
$\alpha^{\dag}_{\lambda}$ and
$\alpha^{\vphantom{\dag}}_{\lambda}$ through the equation:
\be
\left(
\begin{array}{c}
\alpha^{\dag}_{\lambda} \\
\alpha^{\vphantom{\dag}}_{\lambda} \\
\end{array}
\right) = \int dx\,
\left(
\begin{array}{cc}
\psi_{\lambda;+}(x;+)\: & \psi_{\lambda;+}(x;-) \\
\psi_{\lambda;-}(x;+)\: & \psi_{\lambda;-}(x;-) \\
\end{array}
\right)\,
\left(
\begin{array}{r}
a^{\dag} (x) \\
a\,(x) \\
\end{array}
\right)\,.
\label{qpbp}
\ee
With this definition the functions $\psi_{\lambda ; \eta}(x;\chi)$
form the matrix of the Bogoliubov transformation. Then the properties
(\ref{orth}) -- (\ref{conj}) simply follow from the requirement of the
unitarity of this transformation
and from the fact that
$\alpha^{\dag}_{\lambda}$ and
$\alpha^{\vphantom{\dag}}_{\lambda}$
are a Hermitian conjugate pair of the operators.
The inverse relation, which follows from (\ref{comp}) and (\ref{conj}),
reads:
\be
\left(
\begin{array}{r}
a^{\dag} (x) \\
a\,(x) \\
\end{array}
\right) = \sum_{\lambda}
\left(
\begin{array}{cc}
\psi_{\lambda;-}(x;-)\: & \psi_{\lambda;+}(x;-) \\
\psi_{\lambda;-}(x;+)\: & \psi_{\lambda;+}(x;+) \\
\end{array}
\right)\,
\left(
\begin{array}{c}
\alpha^{\dag}_{\lambda} \\
\alpha^{\vphantom{\dag}}_{\lambda} \\
\end{array}
\right)\,.
\label{bpqp}
\ee

Let us define the quasiparticle-vacuum wave function
$\tilde{\Psi}$ by the ordinary condition:
\be
\alpha^{\vphantom{\dag}}_{\lambda}\,
| \tilde{\Psi} \rangle = 0\,,\quad \forall\: \lambda\,.
\label{qpvac}
\ee
This definition is unique, and the wave function $\tilde{\Psi}$
defined by the condition (\ref{qpvac}) is invariant under the
$C$-transformation (\ref{unitr}).
So, one can find the explicit general form of $\tilde{\Psi}$
by using the CB representation (\ref{cbc}).
The result is well known (see, e.~g., Ref.~\cite{RS}) and reads:
\be
| \tilde{\Psi} \rangle = \prod_b
( \tilde{u}^{\vphantom{\dag}}_b +
  \tilde{v}^{\vphantom{\dag}}_b\, a^{\dag}_b )\,\prod_p
( \tilde{u}^{\vphantom{\dag}}_{\vphantom{\bar{p}}p} +
  \tilde{v}^{\vphantom{\dag}}_{\vphantom{\bar{p}}p}\,
a^{\dag}_{\vphantom{\bar{p}}p}\,
a^{\dag}_{\bar{p}} )\,
| \,0\, \rangle \,,
\label{eqpvac}
\ee
where $| \,0\, \rangle$ is the particle-vacuum wave function,
$a^{\dag}_{\lambda} = \int dx\,
\tilde{\phi}^{\vphantom{\dag}}_{\lambda}(x)\,a^{\dag} (x)$.
If, further, we formally define
the EDM $\cal R$ through Eqs.~(\ref{rnot}) with
\be
\tilde{\rho} (x,x') = \langle \tilde{\Psi} |\,
a^{\dag} (x')\, a (x)\, | \tilde{\Psi} \rangle\,,
\label{trho}
\ee
\be
\tilde{\kappa} (x,x') = \langle \tilde{\Psi} |\,
a (x')\, a (x)\, | \tilde{\Psi} \rangle\,,\quad
\tilde{\kappa}^* (x,x') = \langle \tilde{\Psi} |\,
a^{\dag} (x)\, a^{\dag} (x')\, | \tilde{\Psi} \rangle\,,
\label{tkappa}
\ee
then it is easy to show
using Eqs. (\ref{bpqp}) and (\ref{qpvac}) that this EDM
is expressed in terms of the functions $\psi_{\lambda ; \eta}(x;\chi)$
by Eq.~(\ref{defedm}), and consequently it satisfies
Eqs. (\ref{rprop1}) -- (\ref{rprop2}).
In other words, Eqs. (\ref{rnot}), (\ref{trho}), (\ref{tkappa})
set a mapping of $\tilde{\Psi}$ onto the
EDM $\cal R$ satisfying Eqs.~(\ref{defedm}) -- (\ref{rprop2}).
In the following the many-to-one mappings
of the many-fermion wave functions (e.~g., $\tilde{\Psi}$)
to the density matrices (e.~g., $\tilde{\rho}$ or $\cal R$),
defined by equations of the type (\ref{rnot}), (\ref{trho}),
and (\ref{tkappa}), will be denoted as
$\tilde{\Psi} \to \tilde{\rho}$ or $\tilde{\Psi} \to \cal R$.
Due to the fact that in the above derivation the matrix elements of
the Bogoliubov transformation $\psi_{\lambda ; \eta}(x;\chi)$ are
constrained only by conditions (\ref{orth}) -- (\ref{conj}),
the existence of the pointed mapping $\tilde{\Psi} \to \cal R$
is true, in  particular, if $\psi_{\lambda ; \eta}(x;\chi)$
are chosen in such a way that Eqs.~(\ref{rprop3}) are fulfilled
for some arbitrarily given DM $\rho$
(that is always possible as was proved above).
Then we have also the mapping $\tilde{\Psi} \to {\cal R} \to \rho$.
Consequently, it can be argued that for an arbitrary {\sl nonlocal}
DM $\rho (x,x')$, corresponding to some interacting
time-reversal invariant
fermion system, including the case for which
the number of particles is exactly conserved,
there exists a quasiparticle-vacuum wave function $\tilde{\Psi}$
such that $\tilde{\Psi} \to \rho$.
The explicit form of $\tilde{\Psi}$ is given by Eq.~(\ref{eqpvac}),
in which $\tilde{v}^2_{\lambda}$ and
$\tilde{\phi}^{\vphantom{*}}_{\lambda}(x)$
are the eigenvalues ($v^2_{\lambda}$)
and the eigenfunctions ($\phi^{\vphantom{*}}_{\lambda}(x)$)
of the given DM $\rho$. The opposite is also true:
any quasiparticle-vacuum wave function, which satisfies
the condition $\tilde{\Psi} \to \rho$,
has the general explicit form (\ref{eqpvac}) with
$\tilde{u}_{\lambda} = u_{\lambda}$,
$\tilde{v}_{\lambda} = v_{\lambda}$,
$\tilde{\phi}^{\vphantom{*}}_{\lambda}(x) =
\phi^{\vphantom{*}}_{\lambda}(x)$.

These statements can be
considered as a generalization of the Lieb theorem \cite{Lieb}
proved for the local particle density and the Slater-determinant
wave functions. It is remarkable that
the inclusion of the pairing is enough not only to prove a more
general statement but also to make the proof much more simple.
Moreover, the proof of Lieb is based on the particular example
of a Slater-determinant wave function which produces a given
local particle density.
However, in the Lieb theorem the general explicit form of
such Slater-determinant wave function is not constructed.
In contrast, in our case the general explicit form (\ref{eqpvac})
of the quasiparticle-vacuum wave function, which satisfies
the condition $\tilde{\Psi} \to \rho$, is known.

\section{EXTENSION OF THE DFT \label{sec3}}

Let $H$ be the nonrelativistic exact many-body Hamiltonian
of an interacting fermion system. Let us define an auxiliary
functional which depends only on the normal nonlocal DM $\rho$:
\be
E[\rho] = \inf_{\Psi \to \rho}
\left< \Psi \right| H \left| \Psi \right>\,,
\label{df1}
\ee
where $\Psi$ are arbitrary normalized many-fermion wave functions,
including the ones with a {\sl fixed} number of particles.
Following the method of Ref.~\cite{STV} let us introduce
an {\sl effective} many-body Hamiltonian $\tilde{H}$ which
generally does not coincide with $H$. Now we define
\be
\tilde{\cal E}[\rho,\,\tilde{\kappa},\,\tilde{\kappa}^*] =
\inf_{\tilde{\Psi} \to \rho,\,\tilde{\kappa},\,\tilde{\kappa}^*}
\left<\vphantom{\Psi}\right.\tilde{\Psi}
\left.\vphantom{\Psi}\right| \tilde{H}
\left|\vphantom{\Psi}\right.\tilde{\Psi}
\left.\vphantom{\Psi}\right>\,,
\label{df2}
\ee
where $\tilde{\Psi}$ are the quasiparticle-vacuum wave functions.
Due to the existence of the mapping $\tilde{\Psi} \to \rho$
proved in the previous section, the functional
$\tilde{\cal E}[\rho,\,\tilde{\kappa},\,\tilde{\kappa}^*]$
can be defined for an arbitrary nonlocal DM $\rho (x,x')$
corresponding to some interacting
time-reversal invariant fermion system
with a fixed number of particles and for those matrices
$\tilde{\kappa}$, $\tilde{\kappa}^*$ which are produced by the
quasiparticle-vacuum wave functions according to Eqs.~(\ref{tkappa}).
Notice, however, that
Eq.~(\ref{df2}) implies that $\tilde{H}$ is not a completely arbitrary
operator because it is constrained by some mathematical conditions.
First, the energy functional $\tilde{\cal E}$ has to be well defined.
This is not a trivial property
(in spite of the existing mapping $\tilde{\Psi} \to \rho$)
because in the case of atomic nuclei the expectation value
of the exact many-body Hamiltonian $H$ obtained with the
quasiparticle-vacuum wave function can diverge due to the short-range
singularity of the bare nucleon-nucleon ($NN$) forces. Thus, it is
assumed that the effective Hamiltonian $\tilde{H}$ contains $NN$ forces
whose matrix elements are well defined.
Second, $\tilde{H}$ has to be chosen to ensure the minimal property
of the total energy functional
(see Eq.~(\ref{df4}) below and Ref.~\cite{STV} for more details).
Using Eqs.~(\ref{df1}) and (\ref{df2})
we can define, in analogy with Ref.~\cite{STV},
the residual correlation energy $E_{\mbss{RC}}$:
\be
E_{\mbss{RC}}[\rho] = E[\rho] -
\inf_{\tilde{\kappa},\,\tilde{\kappa}^*}
\tilde{\cal E}[\rho,\,\tilde{\kappa},\,\tilde{\kappa}^*]\,,
\label{df3}
\ee
and the total energy functional ${\cal E}$:
\be
{\cal E}[\rho,\,\tilde{\kappa},\,\tilde{\kappa}^*] =
\tilde{\cal E}[\rho,\,\tilde{\kappa},\,\tilde{\kappa}^*] +
E_{\mbss{RC}}[\rho]\,.
\label{df4}
\ee
The main property of the total functional
${\cal E}[\rho,\,\tilde{\kappa},\,\tilde{\kappa}^*]$ is:
\be
\inf_{\rho,\,\tilde{\kappa},\,\tilde{\kappa}^*}
{\cal E}[\rho,\,\tilde{\kappa},\,\tilde{\kappa}^*] =
\inf_{\rho}\left(\inf_{\tilde{\kappa},\,\tilde{\kappa}^*}
\tilde{\cal E}[\rho,\,\tilde{\kappa},\,\tilde{\kappa}^*] +
E_{\mbss{RC}}[\rho] \right) =
\inf_{\rho} E[\rho] = E_{\mbss{GS}}\,,
\label{mprp}
\ee
where $E_{\mbss{GS}}$ is the exact ground-state energy
of the interacting system.
If the infimum of the total functional (\ref{df4}) is the minimum
(usual assumption which has to be fulfilled by an appropriate
choice of the effective Hamiltonian $\tilde{H}$),
it is attained for the {\sl true} nonlocal ground-state
DM $\rho$ as it follows from Eq.~(\ref{df1}) and from the results
of Sec.~\ref{sec2}.

It is advisable to pass from the variables
$\rho$, $\tilde{\kappa}$, and $\tilde{\kappa}^*$
in the functional ${\cal E}$ to the components of the EDM ${\cal R}$
using Eqs.~(\ref{rnot}) for $\tilde{\kappa}$ and $\tilde{\kappa}^*$,
and the relation
\be
\rho (x,x') = \frac{1}{2}\, [\,\delta(x',x)
- {\cal R}(x',-\, ;\, x,-) + {\cal R}(x,+\, ;\, x',+)\,]
\label{rprop7}
\ee
which follows from Eqs.~(\ref{rnot}) at $\tilde{\rho} = \rho$.
Taking into account these relations, we introduce
the energy functional ${\cal E}_{\mbss{ext}}$ depending on the EDM:
\be
{\cal E}_{\mbss{ext}}[{\cal R}] =
{\cal E}[\rho,\,\tilde{\kappa},\,\tilde{\kappa}^*]\,.
\label{defed}
\ee
It is obvious from Eqs.~(\ref{mprp}) and (\ref{defed}) that
$\inf_{\cal R}{\cal E}_{\mbss{ext}}[{\cal R}] = E_{\mbss{GS}}$.

In order to establish the equations of motion of the
theory, which will in the following be referred to as the
Extended Density Matrix Functional Theory (EDMFT),
let us define the functional
\bea
F[\psi^{\vphantom{*}}_{\lambda;-},\,\psi^*_{\lambda;-}] &=&
{\cal E}_{\mbss{ext}}[{\cal R}]
+ \frac{1}{2} \sum_{\lambda, \chi} E_{\lambda}
\int dx\, \left| \psi^{\vphantom{*}}_{\lambda;-}(x;\,\chi) \right|^2
\nonumber\\
&-& \frac{1}{2} \int dx\, dx'\, \mu (x,x')\, [ \delta (x',x) +
\sum_{\chi} \chi\, {\cal R} (x',\,\chi;\,x,\,\chi) ]\,,
\label{deff}
\eea
where $E_{\lambda}$ and $\,\mu (x,x') =  \mu_q\,
\delta (x,x')$ are Lagrange multipliers introduced to ensure the
normalization condition for
$\psi^{\vphantom{*}}_{\lambda;-}(x;\,\chi)$ (see Eq.~(\ref{orth}))
and the neutron and proton numbers conservation:
\be
\sum_{\sigma}
\int d {\mbold r}\, \rho({\mbold r}, \sigma, q;\, {\mbold r},
\sigma, q) = N_q
\label{pnc}
\ee
which are introduced through Eq.~(\ref{rprop7}).
Applying the variational principle to the functional $F$,
we obtain the following set of equations of motion
(see Appendix~A for details):
\be
\sum_{\chi'} \int dx'\, {\cal H} (x,\,\chi;\,x',\,\chi')\,
\psi^{\vphantom{*}}_{\lambda ; \eta}(x';\,\chi') =
\eta\,E_{\lambda}\,\psi^{\vphantom{*}}_{\lambda ; \eta}(x;\,\chi)\,,
\label{eqmex}
\ee
where
\be
{\cal H} (x,\,\chi;\,x',\,\chi') =
2 \frac{\delta \,{\cal E}_{\mbss{ext}}[{\cal R}]}
{\delta \,{\cal R} (x',\,\chi';\,x,\,\chi)}
- \chi\,\delta_{\chi,\chi'}\,\mu (x,x')\,.
\label{defhsp}
\ee
These equations can also be written in the matrix form
\be
\left(
\begin{array}{cc}
h - \mu & \Delta \\ - \Delta^* & \mu -h^* \\
\end{array}
\right)
\left(
\begin{array}{c}
\psi_{\lambda;\eta}^{(+)} \\
\psi_{\lambda;\eta}^{(-)} \\
\end{array}
\right)
= \eta E_{\lambda}
\left(
\begin{array}{c}
\psi_{\lambda;\eta}^{(+)} \\
\psi_{\lambda;\eta}^{(-)} \\
\end{array}
\right)\,,
\label{eqm}
\ee
where $h$ is the single-particle pseudo-Hamiltonian,
$\Delta$ is the operator of the pairing field:
\be
h (x,x') =
\frac{\delta {\cal E}[\rho,\,\tilde{\kappa},\,\tilde{\kappa}^*]}
{\delta \rho (x',x)}\,, \quad
\Delta (x,x') = -2\,
\frac{\delta {\cal E}[\rho,\,\tilde{\kappa},\,\tilde{\kappa}^*]}
{\delta \tilde{\kappa}^* (x',x)}\,,
\label{hddef}
\ee
and it is denoted:
$\psi_{\lambda;\eta}^{(\pm)} = \psi_{\lambda;\eta}(x;\pm)$.
Obviously, we can consider $E_{\lambda} > 0$
(if $E_{\lambda} < 0$, the permutation
$\psi_{\lambda ; +}(x;\chi) \leftrightarrow \psi_{\lambda ; -}(x;\chi)$
is made).
Thus, the Lagrange multipliers $\mu_q$ and $E_{\lambda}$
play the role of
the chemical potential for the nucleons of the type $q$
and the role of
the absolute value of the quasiparticle energy respectively.
From Eq.~(\ref{eqm}) the sense of the indices $\eta$ and $\chi$,
which appear
in the functions $\psi_{\lambda ; \eta}(x;\chi)$ introduced in
Sec.~\ref{sec2}, can be easily understood.
The index $\eta$ is the sign of the eigenvalue in Eq.~(\ref{eqm}),
and $\chi$ denotes the upper and the lower components
of the eigenfunctions.
In most cases one of these components is small, and it completely
vanishes for the blocked states in the CB representation (\ref{cbc})
of the functions $\psi_{\lambda ; \eta}(x;\chi)$.
However, in general, the functions of the CB representation, that is
$\check{\psi}_{\lambda ; \eta}(x;\chi)$, are not solutions
of Eq.~(\ref{eqmex}),
i.~e. they are not eigenfunctions of the operator $\cal H$
(see comments in the Appendix~B).

As it can be seen from Eq.~(\ref{eqm}),
the equations of motion of the EDMFT have the same form
as the ones of the Hartree-Fock-Bogoliubov (HFB) theory.
However, the EDMFT is not reduced to this theory. First of all,
the total energy functional (\ref{df4}) of the EDMFT has a more general
form as compared with the HFB energy functional. Actually, only the part
$\tilde{\cal E}$ corresponds to the HFB functional, whereas the term
$E_{\mbss{RC}}$ has, in general, a more complicated and less obvious
functional dependence on the DM $\rho$.
But the basic difference between the HFB theory and the EDMFT
is determined by the following: the EDMFT is (in principle)
an exact theory in the sense that Eqs.~(\ref{mprp}) are fulfilled.
This is ensured by the fact that the functional $E[\rho]$, which enters
in the term $E_{\mbss{RC}}$, is defined by  Eq.~(\ref{df1}) through a
set of wave functions $\Psi$ which contains the exact wave function
$\Psi_{\mbss{GS}}$ with a {\sl fixed} number of particles.
In this sense one can say that in the EDMFT the number of particles
is exactly conserved, despite the auxiliary quantity $\tilde{\kappa}$,
which has the sense of an anomalous DM
(but which does not coincide with $\kappa_{\mbss{GS}}$),
takes nonzero values. In this context,
the HFB theory can be considered as the
phenomenological realization of the EDMFT.
The relationship between EDMFT and HFB approaches is analogous to
the relationship between the theory developed in Ref.~\cite{STV}
and the density-dependent Hartree-Fock theory, as it was discussed
in more detail in Ref.~\cite{STV}.

\section{REDUCTION TO THE EXTENDED QUASILOCAL THEORY \label{sec4}}

The theory developed in the previous sections is essentially
a nonlocal one as can be seen from Eqs.~(\ref{eqm}) and (\ref{hddef}).
The exact solution of that equations is a rather complicated problem.
However, one can noticeably simplify the task of solving these equations
by reducing the total energy
functional ${\cal E}[\rho,\,\tilde{\kappa},\,\tilde{\kappa}^*]$
to a quasilocal form following the method described in Ref.~\cite{STV}.

In the particular case of atomic nucleus
let us introduce a set of local quantities consisting of
the local particle $n_q$, the kinetic-energy $\tau_q$,
and the spin ${\mbold J}_q$ densities for neutrons and protons
which are obtained from the nonlocal DM $\rho$ as:
\bea
n_q({\mbold r})&=&\sum_{\sigma}\int dx'
\delta (x,x')\rho(x,x')\,,
\label{ql1}\\
\tau _q({\mbold r})&=&\sum_{\sigma}\int dx'
\delta (x,x')({\bnabla}_{\bf r}{\bnabla}_{\bf r'})\rho(x,x')\,,
\label{ql2}\\
{\mbold J}_q({\mbold r})&=&i\sum_{\sigma}\int dx'
\delta({\mbold r}-{\mbold r}') \delta_{q,q'}
[({\mbold \sigma})_{\sigma',\,\sigma}
\times\bnabla_{\bf r}] \rho(x,x')\,.
\label{ql3}
\eea
Note that in contrast to
the analogous definitions of Ref.~\cite{STV}, in this case the exact DM
$\rho$ enters in Eqs. (\ref{ql1}) -- (\ref{ql3})
(see remark after Eq.~(\ref{mprp})), so the quantities
$\tau_{q}$ and ${\mbold J}_{q}$ are the exact (correlated) neutron and
proton kinetic-energy and spin densities respectively.
Explicit expressions for these local quantities, which follow from
Eqs.~(\ref{rprop3}) and from the representation of the EDM in the form
(\ref{defedm}), are:
\bea
n_q({\mbold r}) &=& \sum_{\sigma} \sum_{\lambda}
\left| \psi^{\vphantom{*}}_{\lambda;-}({\mbold r},\sigma,q\,;+)
\right|^2\,,
\label{ql4}\\
\tau_q({\mbold r}) &=& \sum_{\sigma} \sum_{\lambda}
\left| \bnabla\psi^{\vphantom{*}}_{\lambda;-}({\mbold r},\sigma,q\,;+)
\right|^2\,,
\label{ql5}\\
{\mbold J}_q({\mbold r})&=& i \sum_{\sigma,\sigma'} \sum_{\lambda}
\psi^*_{\lambda;-}({\mbold r},\sigma',q\,;+)
[({\mbold \sigma})_{\sigma',\,\sigma}\times{\bnabla}]
\psi^{\vphantom{*}}_{\lambda;-}({\mbold r},\sigma,q\,;+)\,.
\label{ql6}
\eea
We can also define the quantities $\vk_q$ which have the meaning of
local anomalous densities for each kind of nucleons as:
\bea
\vk_q({\mbold r})&=&i\sum_{\sigma}\int dx'
\delta({\mbold r}-{\mbold r}') \delta_{q,q'}
(\sigma_y)_{\sigma',\,\sigma} \tilde{\kappa}(x,x')
\label{ql7}\\
&=& i \sum_{\sigma,\sigma'} \sum_{\lambda}
\psi^*_{\lambda;-}({\mbold r},\sigma',q\,;-)
(\sigma_y)_{\sigma',\,\sigma}
\psi^{\vphantom{*}}_{\lambda;-}({\mbold r},\sigma,q\,;+)\,.
\label{ql8}
\eea

Let us now introduce three kinds of the short notations:
$\rho_{\mbss{QL}} =
\{n_n,n_p,\tau_n,\tau_p,{\mbold J}_n,{\mbold J}_p\}$,
$\vk = \{ \vk_n,\vk_p \}$, and  $\vk^* = \{ \vk^*_n,\vk^*_p \}$.
Using these notations, we define the following
quasilocal energy functionals:
\bea
{\cal E}^{\mbsu{QL1}}
[\rho_{\mbss{QL}},\,\tilde{\kappa},\,\tilde{\kappa}^*] &=&
\inf_{\rho \to \rho_{\mbts{QL}}}
{\cal E}[\rho,\,\tilde{\kappa},\,\tilde{\kappa}^*]\,,
\label{ql9}\\
{\cal E}^{\mbsu{QL2}}[\rho_{\mbss{QL}},\,\vk,\,\vk^*] &=&
\inf_{\rho \to \rho_{\mbts{QL}}}\;
\inf_{\tilde{\kappa} \to \vk}\; \inf_{\tilde{\kappa}^* \to \vk^*}
{\cal E}[\rho,\,\tilde{\kappa},\,\tilde{\kappa}^*]\,.
\label{ql10}
\eea
Note that the mappings
$\rho \to \rho_{\mbss{QL}}$, $\tilde{\kappa} \to \vk$, and
$\tilde{\kappa}^* \to \vk^*$ are established according to
Eqs. (\ref{ql1}) -- (\ref{ql3}) and (\ref{ql7}).
From the definitions (\ref{ql9}) -- (\ref{ql10}),
it immediately follows that the property (\ref{mprp}) is also true
for both energy functionals ${\cal E}^{\mbsu{QL1}}$ and
${\cal E}^{\mbsu{QL2}}$.
Namely, we have
\be
\inf_{\rho_{\mbts{QL}},\,\tilde{\kappa},\,\tilde{\kappa}^*}
{\cal E}^{\mbsu{QL1}}
[\rho_{\mbss{QL}},\,\tilde{\kappa},\,\tilde{\kappa}^*] =
\inf_{\rho_{\mbts{QL}},\,\vk,\,\vk^*}
{\cal E}^{\mbsu{QL2}}[\rho_{\mbss{QL}},\,\vk,\,\vk^*] =
E_{\mbss{GS}}\,.
\label{mprpql}
\ee
In the following we shall refer to the theories based
on the functionals (\ref{ql9}) and (\ref{ql10}) as
the extended quasilocal density functional theories
EQLT1 and EQLT2.

The equations of motion for these quasilocal theories have
the same matrix form (\ref{eqm}),
as for the nonlocal theory, but with different
definitions of the operators $h$ and $\Delta$.
Making use of Eqs. (\ref{hddef}) and (\ref{ql1}) -- (\ref{ql3}),
we obtain
\be
h(x,x') = \delta_{q,q'} \left\{ \left[
-\bnabla_{\bf r} \frac{\hbar^2}{2m_q^*({\mbold r})} \bnabla_{\bf r}
+U_q ({\mbold r}) \right] \delta_{\sigma,\,\sigma'}
- i {\mbold W}_q ({\mbold r})
\cdot [\bnabla_{\bf r} \times {\mbold\sigma}]_{\sigma,\,\sigma'}
\right\}\delta({\mbold r}-{\mbold r}')\,,
\label{hloc}
\ee
where for the quasilocal theory EQLT1:
\be
\frac{\hbar^2}{2m_q^*({\mbold r})} =
\frac{\delta{\cal E}^{\mbsu{QL1}}}{\delta\tau_q({\mbold r})}\,,
\qquad
U_q({\mbold r}) = \frac{\delta{\cal E}^{\mbsu{QL1}}}
{\delta n_q({\mbold r})}\,,
\qquad
{\mbold W}_q ({\mbold r}) = \frac{\delta{\cal E}^{\mbsu{QL1}}}
{\delta{\mbold J}_q({\mbold r})}\,,
\label{muw}
\ee
and analogous expressions with the replacement of
${\cal E}^{\mbsu{QL1}}$ by ${\cal E}^{\mbsu{QL2}}$
in the case of the quasilocal theory  EQLT2. The difference between
these
two versions of the quasilocal approach consists in the
definition of the operator $\Delta$. Within the EQLT1 we
have
\be
\Delta (x,x') = -2\, \frac{\delta {\cal E}^{\mbsu{QL1}}}
{\delta \tilde{\kappa}^* (x',x)}\,,
\label{dnloc}
\ee
as in the nonlocal theory, whereas in the case of the EQLT2
the operator $\;\Delta$ is purely local and it is obtained
using Eq.~(\ref{ql7})
\be
\Delta (x,x') = -2\,i\,
\delta({\mbold r}-{\mbold r}')\, \delta_{q,q'}\,
(\sigma_y)_{\sigma,\,\sigma'}\,
\frac{\delta {\cal E}^{\mbsu{QL2}}}
{\delta \vk^*_q({\mbold r})}\,.
\label{dloc}
\ee

Let us stress that the solution of the equations of motion associated
with both
EQLT1 and EQLT2 enables us (at least in principle) to calculate
the exact values of all the local densities entering in the set
$\rho_{\mbss{QL}}$. The theory of Ref.~\cite{STV} allows
to find the exact values of the local particle densities
$n_q({\mbold r})$ only.
This difference comes from different character
of the many-fermion wave functions used for the building of the
energy functional: a Slater-determinant wave function in \cite{STV}
for which only the Lieb theorem was proved, and a quasiparticle-vacuum
wave function in the present paper which enables one, as it was shown,
to reproduce an arbitrary nonlocal DM $\rho$.

Comparing our approach with theory
developed in Ref.~\cite{OGK}, it should be noted that we do not introduce
any external anomalous pair potential as it was done in \cite{OGK}.
In our method the pairing emerges only as a consequence of the
interaction between the fermions. In absence of the interaction
the pairing field $\Delta$ vanishes, which is in agreement with the
particle-number conservation condition. The same difference
exists between the equations of motion of the EQLT2 case and
the Bogoliubov-de~Gennes equations \cite{dG} in which the pair
potential enters, in fact, as an external field (though created
initially by the interaction).

It should also be pointed out that the EDMFT in its nonlocal and
quasilocal versions can be reduced to the more simple DFT plus BCS
description of the pairing correlations if the operator $\cal H$ in
Eq.~(\ref{eqmex}) is (or is assumed to be) diagonal in the CB
representation (\ref{cbc}). However, in general, this description is not
equivalent to the one obtained by solving exactly the EDMFT equations of
motion (see comments in the Appendix~B).
This reduction of the EDMFT is analogous to the replacement of the HFB
equations by coupled Hartree-Fock plus BCS equations how it is discussed
in Refs.~\cite{RS,edm}.

\section{SUMMARY AND CONCLUSIONS}

We have extended the recently developed quasilocal density functional theory
\cite{STV} to include pairing correlations. This new approach named
Extended Density Matrix Functional Theory
is based on an extended density matrix (EDM) formalism.
The EDM $\cal R$ contains as a block the normal
density matrix $\rho$ and as another block an auxiliary quantity
$\tilde{\kappa}$ which has the sense of an anomalous density matrix, but which
does not coincide with the exact one $\kappa$ in general. The matrix
$\tilde{\kappa}$ is chosen in such a way that the equality
${\cal R}^2 = {\cal R}$ is fulfilled.
The EDM which possesses this property
can be easily constructed for a given density matrix  $\rho$
using the canonical basis in which $\rho$ is diagonal.
It has been shown that for an arbitrary density
matrix $\rho$, corresponding to some interacting
time-reversal invariant fermion system, there exist a
qusiparticle-vacuum wave function $\tilde{\Psi}$ and an EDM
$\cal R$ such that the following many-to-one mappings take place:
$\tilde{\Psi} \to {\cal R} \to \rho$. This statement
can be considered as a generalization
of the Lieb theorem.
Using the connection between $\cal R$ and $\rho$,
we have defined the total energy functional as an
extended functional of $\cal R$.
It has been proved that its minimum value is equal to
the exact ground state energy of the considered system.
This extended energy functional is
reduced further to a quasilocal form.
Thus, in the corresponding final equations of motion the
single-particle pseudo-Hamiltonian $h$
and the pairing field $\Delta$ are both (quasi) local.
Although the equations of motion in our theory have the same form as the
ones of the Hartree-Fock-Bogoliubov theory, the main difference is that
the Extended Density Matrix Functional Theory is,
in principle, an exact theory in the sense that the
true ground stated energy can be reached
for the true normal density matrix $\rho$.
Finally, notice that the general formalism developed in this work,
which introduces the pairing correlations in the DFT,
has been discussed in the particular case of the atomic nucleus,
but it can be easily applied to other Fermi systems
with a fixed number of particles.

\vspace{3em}
V.~B.~S. and V.~I.~T. would like to acknowledge financial support from the
Russian Ministry of Education under grant No. E02-3.3-463.
V.~I.~T. thanks the Institut f\"ur
Kernphysik at the Forschungszentrum J\"ulich for its hospitality
and financial support during the completion of this work.
X.~V. acknowledges financial support from DGI and FEDER (Spain)
under grant No. BFM2002-01868 and from DGR (Catalonia) under grant
No. 2001SGR-00064.
\vspace{3em}

\begin{flushleft}
{\normalsize\bf APPENDIX A}
\end{flushleft}
In this Appendix the equations of motion of the EDMFT are derived.
Varying the functional $F$ defined by Eq.~(\ref{deff})
and taking into account Eq.~(\ref{defedm}), we obtain:
$$
\begin{array}{rcl}
\delta F &=&
\ds\sum_{\lambda,\chi,\chi'} \int dx\,dx'\left\{
\frac{\delta \,{\cal E}_{\mbss{ext}}[{\cal R}]}
{\delta \,{\cal R} (x',\,\chi';\,x,\,\chi)}
- \frac{1}{2}\,\delta_{\chi,\chi'} \left[\,\chi\,\mu (x,x')
- \delta (x,x')\,E_{\lambda}\,\right]\right\}\\
&\times& \left[\,\psi^{\vphantom{*}}_{\lambda ; -}(x';\,\chi')
\,\delta \psi^*_{\lambda;-}(x;\chi)
+ \psi^*_{\lambda;-}(x;\chi)\,
\delta\psi^{\vphantom{*}}_{\lambda ; -}(x';\,\chi')\, \right] = 0\,.\\
\vphantom{\delta F} &\vphantom{=}&
\vphantom{
\ds\sum_{\lambda,\chi,\chi'} \int dx\,dx'\left\{
\frac{\delta \,{\cal E}_{\mbss{ext}}[{\cal R}]}
{\delta \,{\cal R} (x',\,\chi';\,x,\,\chi)}
- \frac{1}{2}\,\delta_{\chi,\chi'} \left[\,\chi\,\mu (x,x')
- \delta (x,x')\,E_{\lambda}\,\right]\right\}
}\\
\end{array}
\eqno (\mbox{A1})
$$
\vspace{-3.5em}

\noindent
This leads to the following equations of motion:
$$
\sum_{\chi'} \int dx'\, {\cal H} (x,\,\chi;\,x',\,\chi')\,
\psi^{\vphantom{*}}_{\lambda ; -}(x';\,\chi') =
- E_{\lambda}\,\psi^{\vphantom{*}}_{\lambda ; -}(x;\,\chi)\,,
\eqno (\mbox{A2})
$$
where the operator ${\cal H}$ is defined by Eq.~(\ref{defhsp}).
The Eq.~(A2) formally defines only
one half of the complete set of the eigenfunctions of ${\cal H}$.
In order to define second half, let us note first
that if the following equalities are fulfilled
$$
{\cal H} (x,\,\chi;\,x',\,\chi') =
- {\cal H} (x',-\chi';\,x,-\chi)
= {\cal H}^* (x',\,\chi';\,x,\,\chi)\,,
\eqno (\mbox{A3})
$$
then the functions
$\psi^{\vphantom{*}}_{\lambda ; +}(x;\,\chi)$,
defined through the functions
$\psi^{\vphantom{*}}_{\lambda ; -}(x;\,\chi)$
using the condition (\ref{conj}), are also the
eigenfunctions of ${\cal H}$. In this case the complete set
of the eigenfunctions of ${\cal H}$ is divided into two equal parts
with eigenvalues $+E_{\lambda}$ and $-E_{\lambda}$.
On the other hand, if the condition (\ref{conj}) is fulfilled,
and if the set of the functions
$\psi^{\vphantom{*}}_{\lambda ; -}(x;\,\chi)$
is taken as a half of the complete set
$\{ \psi^{\vphantom{*}}_{\lambda ; \pm}(x;\,\chi) \}$,
then Eq.~(\ref{rprop2}) is true, and it follows from Eq.~(\ref{defhsp})
that Eqs.~(A3) are also fulfilled. Consequently, properties
(\ref{conj}) and (A3) follow from each other,
and there exists a solution of the equations of motion
which possesses both properties.
So, setting Eq.~(\ref{conj}) to be satisfied,
we actually choose a solution which has the symmetry
defined by this equation
without imposing additional constraints in the variational procedure.
In that case, using Eqs.~(\ref{conj}) and (A3),
we come from (A2) to Eq.~(\ref{eqmex}) which represents
complete set of the equations of motion.

\begin{flushleft}
{\normalsize\bf APPENDIX B}
\end{flushleft}
In this Appendix the relationship between the eigenfunctions
of the operators $\cal R$ and $\cal H$ is analyzed.
First of all, notice that from Eqs. (\ref{defedm}) and
(\ref{orth}) it follows that
$$
\sum_{\chi'} \int dx'\, {\cal R} (x,\,\chi;\,x',\,\chi')\,
\psi^{\vphantom{*}}_{\lambda ; \eta}(x';\,\chi') =
\delta_{\eta\,, -}\,
\psi^{\vphantom{*}}_{\lambda ; \eta}(x;\,\chi)\,.
\eqno (\mbox{B1})
$$
Thus, the functions $\psi_{\lambda ; \eta}(x;\chi)$ are
eigenfunctions of the operator $\cal R$. Obviously, the set
of the eigenfunctions of $\cal R$ is determined by a given $\cal R$
up to an arbitrary unitary transformation
of the type Eq.~(\ref{unitr}). Consequently, the functions
$\check{\psi}_{\lambda ; \eta}(x;\chi)$ defined by
Eqs.~(\ref{cbc}) are also eigenfunctions of the operator $\cal R$.
On the other hand, from Eqs. (B1) and (\ref{eqmex})
it follows that there exists at least one set of eigenfunctions
(namely the set $\{ \psi_{\lambda ; \eta}(x;\chi) \}$)
which is common for both operators $\cal R$ and $\cal H$.
However, from these equations
it does not follow that any eigenfunction of $\cal R$,
and in particular $\check{\psi}_{\lambda ; \eta}(x;\chi)$,
will be also eigenfunction of $\cal H$.
Indeed, while a set of eigenfunctions of
$\cal R$ is determined up to an arbitrary unitary transformation
(\ref{unitr}), this is not true for the eigenfunctions of
$\cal H$, since Eqs.~(\ref{eqmex}) and (\ref{eqm}) are not covariant
under this transformation.
Consequently, if we use an arbitrarily given
representation of the eigenfunctions
of the operator $\cal R$, in particular the CB representation (\ref{cbc}),
an additional nontrivial $C$-transformation, which does not change
$\cal R$, is generally required to diagonalize $\cal H$
(see Refs.~\cite{RS,edm} for details).
\newpage
\end{document}